\documentclass[11pt,a4paper]{article}

\usepackage{jcappub,amsfonts,amsmath,amssymb,bm,graphicx}

\usepackage[active]{srcltx}

\title{Neutrino spin oscillations in matter under the influence of gravitational
and electromagnetic fields}

\author[a,b]{Maxim Dvornikov}
\affiliation[a]{Institute of Physics, University of S\~{a}o Paulo, CP 66318, CEP 05315-970 S\~{a}o Paulo, SP, Brazil}
\affiliation[b]{Pushkov Institute of Terrestrial Magnetism,
Ionosphere and Radiowave Propagation (IZMIRAN), 142190 Moscow, Troitsk, Russia}
\emailAdd{maxim.dvornikov@usp.br}

\abstract{
We derive the new quasi-classical equation for the description of
the spin evolution of a neutrino propagating in a curved space-time
and interacting with a background matter and an external electromagnetic
field. This equation is used to analyze neutrino spin oscillations
in these external backgrounds. We obtain the effective Hamiltonian
and the transition probability for oscillations of neutrinos when they
move in the vicinity of a rotating black hole, surrounded by an accretion
disk, and interact with an external magnetic field. The appearance
of new resonances in neutrino spin oscillations in this system is considered.
The approximate treatment of spin oscillations of radially propagating ultra high energy neutrinos is developed.
We also discuss the applications of our results to the description
of neutrino spin oscillations in realistic astrophysical media.
}

\keywords{neutrino theory, GR black holes, magnetic fields, accretion}

\begin{document}

\maketitle

\section{Introduction}

Neutrinos is an important ingredient of various astrophysical and
cosmological media~\cite{GiuKim07}. Despite their weak interaction
with other particles, neutrinos can significantly affect the last
stages of the evolution of massive stars. If the mass of a star is
$\sim(10-25)$ solar masses, it can form a neutron star through a
core-collapsing supernova stage with the emission of $99\ \%$ of
the initial gravitational energy in the form of neutrinos~\cite{Heg03}.
Such a dense neutrino flux is sure to influence the dynamics of the
collapse~\cite{Jan13}. A more massive star can become a black hole (BH). A space-time
around BH has an essentially nontrivial geometry leading to the generation
of a strong gravitational field, which, in its turn, can significantly
affect the behavior of emitted neutrinos.

For the first time the influence of a spherically symmetric gravitational
field on the propagation of massless neutrinos, described in frames
of the Lee and Yang theory, was considered in ref.~\cite{BriWhe57}.
Since then a great number of works on this subject, mainly in the
context of neutrino oscillations, have been published. Firstly, we
mention ref.~\cite{CaiPap91}, where the effect of a weak gravitational
field was accounted for in the phase of the neutrino wave function.
On the basis of this approach the neutrino spin-flip owing to the gravity
interaction was considered. The similar idea was used in refs.~\cite{PirRoyWud96,CarFul97}
to include various neutrino mass eigenstates, the interaction with a
background matter and an external electromagnetic field. The evolution
of a single neutrino mass eigenstate in presence of a gravitational
field was described by the classical Hamilton-Jacobi equation in refs.~\cite{PirRoyWud96,CarFul97}. This
approach was recently applied in ref.~\cite{CapLauVer10} to study neutrino
oscillations in frames of the modified theory of gravity.

Although it was experimentally proven that there are several neutrino
mass eigenstates and they are mixed (see, e.g., ref.~\cite{Ahn12}),
still the problem of the neutrino spin-flip within one mass eigenstate,
or neutrino spin oscillations, is of importance. The change of the
neutrino helicity, which converts an active left-polarized neutrino
into its sterile right-polarized counterpart, can be observable if other
channels of neutrino oscillations are suppressed.

A quasi-classical approach for the description of the evolution of
a spinning particle in a gravitational field, different form that
proposed in ref.~\cite{CaiPap91}, was developed in ref.~\cite{Pap51}.
The basic equations of this approach were modified in ref.~\cite{PomKhr98}
to exclude the influence of the spin on the particle trajectory.
The basic equations derived in ref.~\cite{PomKhr98} were applied in
refs.~\cite{Dvo06,AlaHos13} for the analysis of neutrino spin oscillations
in gravitational fields of various configurations.

In the present work we continue our previous studies of gravity induced
neutrino spin oscillations. We generalize the basic equation for the
description of the neutrino spin evolution derived in ref.~\cite{Dvo06}
to include the neutrino interaction with an external electromagnetic
field and a background matter when a particle moves in a curved space-time.
Although a neutrino is an electrically neutral particle, it may interact
with an electromagnetic field by its anomalous magnetic moments~\cite{BroGiuStu12}.
Note that the quasi-classical approach for the description of neutrino
spin oscillations in matter and an electromagnetic field in the flat
space-time was developed in refs.\cite{DvoStu02,LobStu01}.

This paper is organized as follows. In section~\ref{sec:GENFORM} we
generalize the covariant equation for the neutrino spin evolution
in a curved space-time in a way to include the neutrino interaction
with background matter. We also rewrite this equation in the more
convenient form for the three vector of the neutrino spin. In section~\ref{sec:KERR}
we discuss the neutrino motion in the gravitational field of a rotating
BH surrounded by an accretion disk and embedded into a magnetic field.
First, we consider the neutrino motion around BH on a circular orbit with an arbitrary radius and derive the transition
probability for spin oscillations in this case. Then, we study ultra high energy (UHE) neutrinos which propagate almost radially from BH. The approximate effective Hamiltonian for spin oscillations of these particles is derived. We analyze the dynamics
of spin oscillations for several particular neutrino trajectories in section~\ref{sec:APPL}.
Finally, in section~\ref{sec:CONCL}, we briefly summarize our results.

\section{Covariant description of the neutrino spin evolution\label{sec:GENFORM}}

In this section we shall construct the quasi-classical approach for
the description of the spin evolution of a Dirac neutrino moving in
a background matter under the influence of electromagnetic and gravitational
fields. In particular, we generalize the neutrino spin evolution equation
in matter and electromagnetic fields to include the effects of the nontrivial
geometry of space-time.

Let us first remind the basic description of the neutrino spin evolution
when a particle propagates in the flat space-time and interacts with
a background matter and an external electromagnetic field $F_{\mu\nu}$.
Assuming the forward scattering approximation, the covariant equation
for the four vector of the neutrino spin, $S^{\mu}$, has the form~\cite{DvoStu02},
\begin{equation}\label{eq:BMTMink}
  \frac{\mathrm{d}S^{\mu}}{\mathrm{d}\tau}=
  2\mu
  \left(
    F^{\mu\nu}S_{\nu}-U^{\mu}U_{\nu}F^{\nu\lambda}S_{\lambda}
  \right) +
  \sqrt{2}G_{\mathrm{F}}\varepsilon^{\mu\nu\lambda\rho}G_{\nu}U_{\lambda}S_{\rho},
\end{equation}
where $U^{\mu}$ is the neutrino four velocity, $\tau$ is the interval,
$\mu$ is the neutrino magnetic moment, $\varepsilon^{\mu\nu\lambda\rho}$
is the absolute antisymmetric tensor in the Minkowski space with $\varepsilon^{0123}=1$,
and $G_{\mathrm{F}}$ is the Fermi constant. In this approximation
the neutrino four velocity is constant.

The interaction with a background matter is characterized by the four
vector $G^{\mu}$, which is a linear combination of hydrodynamic currents,
$J_{f}^{\mu}$, and polarizations, $\Lambda_{f}^{\mu}$, of background
fermions of the type $f$,
\begin{equation}\label{eq:Gmugen}
  G^{\mu} = \sum_{f}
  \left(
    q_{f}^{(1)}J_{f}^{\mu}+q_{f}^{(2)}\Lambda_{f}^{\mu}
  \right).
\end{equation}
We shall express $J_{f}^{\mu}$ using the invariant number density
(i.e. the density in the rest frame of fermions), $n_{f}$, and four velocity,
$U_{f}^{\mu}$, of background fermions: $J_{f}^{\mu}=n_{f}U_{f}^{\mu}$.
The explicit expression of $\Lambda_{f}^{\mu}$ is given in ref.~\cite{LobStu01}.
The coefficients $q_{f}^{(1,2)}$ in eq.~(\ref{eq:Gmugen}) were
found in ref.~\cite{DvoStu02}. In case we study the propagation of
$\nu_{e}$ in matter consisting of electrons, protons, and neutrons
these coefficients take the form,
\begin{equation}\label{eq:q1q2nue}
  q_{1}^{(f)} = I_{\mathrm{L}3}^{(f)}-2Q_{f}\sin^{2}\theta_{\mathrm{W}}+\delta_{ef},
  \quad
  q_{2}^{(f)}=-I_{\mathrm{L}3}^{(f)}-\delta_{ef},
\end{equation}
where $I_{\mathrm{L}3}^{(f)}$ is the third component of the weak
isospin of type $f$ fermions, $Q_{f}$ is the value of their electric
charge, $\theta_{\mathrm{W}}$ is the Weinberg angle, $\delta_{ef}=1$
for electrons and vanishes for protons and neutrons. To get these
coefficients for $\nu_{\mu,\tau}$ interactions with the same background
fermions, we should put the symbol $\delta_{ef}$ to zero in eq.~(\ref{eq:q1q2nue}).

The dynamics of the spin of an elementary particle in presence of
a gravitational field was studied in ref.~\cite{PomKhr98}. To account
for the influence of a nontrivial space-time geometry
one should replace $\mathrm{d}/\mathrm{d}\tau\to\mathrm{D}/\mathrm{D}\tau$ in eq.~(\ref{eq:BMTMink}),
where $\mathrm{D}/\mathrm{D}\tau$ is the covariant derivative along
the world line. The evolution of a spin associated with the magnetic
moment interaction with an external electromagnetic field in a curved
space-time was studied in ref.~\cite{BalKurZim02}.

The most straightforward generalization of the contribution of the
neutrino-matter interaction in presence of an external gravitational
field consists in the replacement $\varepsilon^{\mu\nu\lambda\rho}\to E^{\mu\nu\lambda\rho}$
in eq.~(\ref{eq:BMTMink}), where $E^{\mu\nu\lambda\rho}=\tfrac{1}{\sqrt{-g}}\varepsilon^{\mu\nu\lambda\rho}$
is the covariant antisymmetric tensor in a curved space-time and $g=\det(g_{\mu\nu})$
is the determinant of the metric tensor $g_{\mu\nu}$. Eventually
we get the evolution equation for the spin of a neutrino propagating
in a background matter and interacting with external electromagnetic
and gravitational fields,
\begin{equation}\label{eq:BMTcurvedst}
  \frac{\mathrm{D}S^{\mu}}{\mathrm{D}\tau} =
  2\mu
  \left(
    F^{\mu\nu}S_{\nu}-U^{\mu}U_{\nu}F^{\nu\lambda}S_{\lambda}
  \right) +
  \sqrt{2}G_{\mathrm{F}}E^{\mu\nu\lambda\rho}G_{\nu}U_{\lambda}S_{\rho}.
\end{equation}
We should also remind that, despite we use the forward scattering
approximation, the trajectory of a neutrino is no longer a straight
line while a particle moves in a curved space-time. To account for
the deflection of a trajectory one should consider the geodesic equation for
the evolution of the four velocity,
\begin{equation}\label{eq:Trajcurvedst}
  \frac{\mathrm{D}U^{\mu}}{\mathrm{D}\tau} = 0,
\end{equation}
along with eq.~(\ref{eq:BMTcurvedst}). The validity of eq.~\eqref{eq:Trajcurvedst} will be discussed in section~\ref{sec:CONCL}.

To describe the possible polarization of background fermions in eqs.~\eqref{eq:Gmugen} and~(\ref{eq:BMTcurvedst})
we introduce a locally Minkowskian frame,
\begin{equation}\label{eq:polarizvierb}
  \lambda_{f}^{a} = e_{\ \mu}^{a}\Lambda_{f}^{\mu},
\end{equation}
where $e_{\ \mu}^{a}$, $a=0,\dots,3$, are the vierbein vectors satisfying
the relations
\begin{equation}\label{eq:vierbdef}
  g_{\mu\nu} = e_{\ \mu}^{a}e_{\ \nu}^{b}\eta_{ab},
  \quad
  \eta_{ab} = e_{a}^{\ \mu}e_{a}^{\ \nu}g_{\mu\nu}.
\end{equation}
Here $e_{a}^{\ \mu}$ is the inverse vierbein ($e_{a}^{\ \mu}e_{\ \nu}^{a}=\delta_{\nu}^{\mu}$
and $e_{a}^{\ \mu}e_{\ \mu}^{b}=\delta_{a}^{b}$) and $\eta_{ab}=\mathrm{diag}(1,-1,-1,-1)$
is the metric tensor in the Minkowski space. The fermions polarization
in the vierbein frame, $\lambda_{f}^{a}$, is related to the invariant
fermion polarization (the polarization in the frame where fermions are at
rest), which is a three vector, $\bm{\zeta}_{f}$, by
\begin{equation}\label{eq:polvierbzeta}
  \lambda_{f}^{a} = n_{f}
  \left(
    (\bm{\zeta}_{f}\mathbf{u}_{f}),
    \bm{\zeta}_{f}+\frac{\mathbf{u}_{f}(\bm{\zeta}_{f}\mathbf{u}_{f})}{1+u_{f}^{0}}
  \right),
\end{equation}
where $u_{f}^{a}=(u_{f}^{0},\mathbf{u}_{f})=e_{\ \mu}^{a}U_{f}^{\mu}$,
cf. eq.~(\ref{eq:polarizvierb}), is the four velocity of background
fermions in the vierbein frame.

Although eqs.~(\ref{eq:BMTcurvedst}) and~(\ref{eq:Trajcurvedst})
can be used to describe the evolution of the neutrino spin, it is
more convenient to rewrite them in the vierbein frame introducing
$s^{a}=e_{\ \mu}^{a}S^{\mu}$ and $u^{a}=e_{\ \mu}^{a}U^{\mu}=(u^{0},\mathbf{u})$.
Using eq.~(\ref{eq:BMTcurvedst}) we get the evolution equation for
$s^{a}$ in the following form:
\begin{equation}\label{eq:BMTvierb}
  \frac{\mathrm{d}s^{a}}{\mathrm{d}t} =
  \frac{1}{\gamma}
  \left[
    G^{ab}s_{b}+2\mu
    \left(
      f^{ab}s_{b}-u^{a}u_{b}f^{bc}s_{c}
    \right) +
    \sqrt{2}G_{\mathrm{F}}\varepsilon^{abcd}g_{b}u_{c}s_{d}
  \right],
\end{equation}
where $G^{ab}=\eta^{ac}\eta^{bd}\gamma_{cde}u^{e}$ is the antisymmetric
tensor accounting for the gravitational interaction of neutrinos,
$\gamma_{abc}=\eta_{ad}e_{\ \mu;\nu}^{d}e_{b}^{\ \mu}e_{c}^{\ \nu}$
are the Ricci rotation coefficients, the semicolon
stays for a covariant derivative, $\gamma=U^{0}$, $g^{a}=e_{\ \mu}^{a}G^{\mu}=(g{}^{0},\mathbf{g})$
is the effective potential of the matter interaction in the vierbein
frame, and $f_{ab}=e_{a}^{\ \mu}e_{b}^{\ \nu}F_{\mu\nu}=(\mathbf{e},\mathbf{b})$
is the electromagnetic field tensor in the vierbein frame. To derive
eq.~(\ref{eq:BMTvierb}) we use the fact that $\varepsilon^{abcd} = e_{\ \mu}^{a}e_{\ \nu}^{b}e_{\ \lambda}^{c}e_{\ \rho}^{d}E^{\mu\nu\lambda\rho}$.

We can also reformulate the dynamics of $U^{\mu}$ in the vierbein frame.
One can obtain the following evolution equation for $u^{a}$:
\begin{equation}\label{eq:Uvierb}
  \frac{\mathrm{d}u^{a}}{\mathrm{d}t} = \frac{1}{\gamma}G^{ab}u_{b}.
\end{equation}
The details of the derivation of eqs.~(\ref{eq:BMTvierb}) and~(\ref{eq:Uvierb})
can be found in refs.~\cite{PomKhr98,Dvo06}.

Analogously to eq.~(\ref{eq:polvierbzeta}) we express $s^{a}$ in
terms of the three vector of the neutrino spin, $\bm{\zeta}$,
\begin{equation}\label{eq:nuspinzeta}
  s^{a} =
  \left(
    (\bm{\zeta}\mathbf{u}),
    \bm{\zeta}+\frac{\mathbf{u}(\bm{\zeta}\mathbf{u})}{1+u^{0}}
  \right).
\end{equation}
Using
eqs.~(\ref{eq:BMTvierb}) and~(\ref{eq:nuspinzeta}) one finds the
evolution equation for $\bm{\zeta}$ as
\begin{equation}\label{eq:nuspinrot}
  \frac{\mathrm{d}\bm{\zeta}}{\mathrm{d}t} = \frac{2}{\gamma}[\bm{\zeta}\times\mathbf{G}],
\end{equation}
where
\begin{align}\label{eq:vectG}
  \mathbf{G} = &
  \frac{1}{2}
  \left[
    \mathbf{b}_{g}+\frac{1}{1+u^{0}}
    \left(
      \mathbf{e}_{g}\times\mathbf{u}
    \right)
  \right] +
  \frac{G_{\mathrm{F}}}{\sqrt{2}}
  \left[
    \mathbf{u}
    \left(
      g^{0}-\frac{(\mathbf{g}\mathbf{u})}{1+u^{0}}
    \right) -
    \mathbf{g}
  \right]
  \notag
  \\
  & +
  \mu
  \left[
    u^{0}\mathbf{b}-\frac{\mathbf{u}(\mathbf{u}\mathbf{b})}{1+u^{0}}+(\mathbf{e}\times\mathbf{u})
  \right].
\end{align}
Here $\mathbf{e}_{g}$ and $\mathbf{b}_{g}$ are the components of
the tensor $G_{ab}$: $G_{ab}=(\mathbf{e}_{g},\mathbf{b}_{g})$. To
derive eq.~(\ref{eq:vectG}) we use the fact that $u_{a}u^{a}=1$.

\section{Neutrino propagation in the vicinity of a rotating black hole\label{sec:KERR}}

In this section we describe the dynamics of neutrino spin oscillations
in matter and magnetic field when a particle moves in the vicinity
of a rotating BH. First, we shall consider circular neutrino motion and derive the effective Schr\"{o}dinger equation
as well as find the transition probability for spin oscillations. Then, we shall analyze the radial neutrino propagation and obtain the approximate effective Hamiltonian in this case.

The space-time of a rotating BH can be described by the Kerr metric.
In Boyer-Lindquist coordinates $x^\mu = (t,r,\theta,\phi)$ this metric has the form,
\begin{equation}\label{eq:Kerrmetr}
  \mathrm{d}\tau^{2} =
  \left(
    1-\frac{rr_{g}}{\Sigma}
  \right)
  \mathrm{d}t^{2} +
  2\frac{rr_{g}a\sin^{2}\theta}{\Sigma}\mathrm{d}t\mathrm{d}\phi -
  \frac{\Sigma}{\Delta}\mathrm{d}r^{2} -
  \Sigma\mathrm{d}\theta^{2} -
  \frac{\Xi}{\Sigma}\sin^{2}\theta\mathrm{d}\phi^{2},
\end{equation}
where
\begin{equation}\label{eq:metrpar}
  \Delta=r^{2}-rr_{g}+a^{2},
  \quad
  \Sigma=r^{2}+a^{2}\cos^{2}\theta,
  \quad
  \Xi=\left(r^{2}+a^{2}\right)\Sigma+rr_{g}a^{2}\sin^{2}\theta.
\end{equation}
We use units where the gravitational constant is equal to one. In
this case the mass of BH $M=r_{g}/2$ and its angular momentum $J=Ma$.
Coordinates $x^\mu$ in eqs.~\eqref{eq:Kerrmetr} and~\eqref{eq:metrpar} correspond to the world frame rather than to the vierbein frame.

One can check by means of the direct calculation that the following
vectors:
\begin{align}
  e_{0}^{\ \mu}= &
  \left(
    \sqrt{\frac{\Xi}{\Sigma\Delta}},0,0,\frac{arr_{g}}{\sqrt{\Delta\Sigma\Xi}}
  \right),
  \nonumber \\
  e_{1}^{\ \mu}= &
  \left(
    0,\sqrt{\frac{\Delta}{\Sigma}},0,0
  \right),
  \nonumber \\
  e_{2}^{\ \mu}= &
  \left(
    0,0,\frac{1}{\sqrt{\Sigma}},0
  \right),
  \nonumber \\
  e_{3}^{\ \mu}= &
  \left(
    0,0,0,\frac{1}{\sin\theta}\sqrt{\frac{\Sigma}{\Xi}}
  \right),
  \label{eq:vierbKerr}
\end{align}
satisfy the condition in eq.~(\ref{eq:vierbdef}). Thus they can be taken as
vierbein vectors. On the basis of eq.~\eqref{eq:vierbKerr}, one can reproduce the vierbein vectors used in ref.~\cite{Dvo06}. Those vectors correspond to $a=0$, i.e. to a nonrotating BH.

We shall discuss the case of BH surrounded by matter forming an accretion
disk. In this case only $U_{f}^{0}$ and $U_{f}^{\phi}$ are nonzero.
Moreover we shall consider the stationary accretion. By the symmetry
reasons, the quantities $n_{f}$, $U_{f}^{0}$, and $U_{f}^{\phi}$
can be functions of $r$ and $\theta$ only.

Now let us specify the structure of the electromagnetic field. We
suggest that at relatively great distance from BH there is a constant
and uniform magnetic field parallel to the rotation axis of BH and
having the strength $B_{0}$. This magnetic field can be created by the
plasma motion in the accretion disk. The structure of the axially symmetric
electromagnetic field in a curved space-time having such an asymptotics
was studied in ref.~\cite{Wal74}. The nonzero components of the
vector potential of such a field have the form,
\begin{equation}\label{eq:vectpot}
  A_{t} =
  B_{0}a
  \left[
    1-\frac{rr_{g}}{2\Sigma}(1+\cos^{2}\theta)
  \right],
  \quad
  A_{\phi} =
  -\frac{B_{0}}{2}
  \left[
    r^{2}+a^{2}-\frac{rr_{g}a^{2}}{\Sigma}(1+\cos^{2}\theta)
  \right]
  \sin^{2}\theta.
\end{equation}
The electric and magnetic field strengths can be calculated on the
basis of eq.~(\ref{eq:vectpot}) as $F_{\mu\nu}=\partial_{\mu}A_{\nu}-\partial_{\nu}A_{\mu}$.
Note that we do not study the modification of the metric in eq.~(\ref{eq:Kerrmetr})
by the external electromagnetic field, as it was considered in ref.~\cite{Ern76}.

Using eqs.~(\ref{eq:vectG}) and~(\ref{eq:vierbKerr}) we can describe
the neutrino spin evolution for an arbitrary neutrino trajectory in
the gravitational field of a rotating BH. However to proceed in analytical
analysis first we shall study the circular neutrino motion
in the equatorial plane, i.e. when $U^{\theta}=0$, $\theta=\pi/2$,
and $U^{r}=0$. Then the components of the four velocity have the
form,
\begin{equation}\label{eq:uKerr}
  u^{a} =
  \left(
    rU^{0}\sqrt{\frac{\Delta}{\Xi_{0}}},0,0,\frac{U^{\phi}\Xi_{0}-arr_{g}U^{0}}{r\sqrt{\Xi_{0}}}
  \right),
\end{equation}
where $\Xi_{0}=\Xi(\theta=\pi/2)=r^{4}+a^{2}r(r+r_{g})$.

Just for simplicity we shall consider that background matter is unpolarized.
It is the case when the temperature of background fermions is sufficiently
high. Analogously to eq.~(\ref{eq:uKerr}) we get that the neutrino
interaction with background fermions is described by the four vector,
\begin{equation}\label{eq:gKerr}
  g^{a} =
  n_{\mathrm{eff}}
  \left(
    rU_{f}^{0}\sqrt{\frac{\Delta}{\Xi_{0}}},0,0,\frac{U_{f}^{\phi}\Xi_{0}-arr_{g}U_{f}^{0}}{r\sqrt{\Xi_{0}}}
  \right),
\end{equation}
where
\begin{equation}\label{eq:neff}
  n_{\mathrm{eff}} = \sum_{f}q_{f}^{(1)}n_{f},
\end{equation}
is the effective number density. In eqs.~(\ref{eq:gKerr}) and~(\ref{eq:neff})
we assume that all types of background fermions have the same
four velocity.

Using eq.~(\ref{eq:vierbKerr}) and the formalism developed in ref.~\cite{Dvo06},
we get the nonzero components of the vectors $\mathbf{e}_{g}$ and
$\mathbf{b}_{g}$ as
\begin{align}\label{eq:egbgKerr}
  e_{g1} = &
  -\frac{r_{g}}{2r^{2}\sqrt{\Xi_{0}}}
  \left[
    U^{0}(r^{2}+a^{2})-aU^{\phi}(3r^{2}+a^{2})
  \right],
  \notag
  \\
  b_{g2} = &
  -\frac{1}{2r^{2}}\sqrt{\frac{\Delta}{\Xi_{0}}}
  \left[
    U^{\phi}(2r^{3}-a^{2}r_{g})+ar_{g}U^{0}
  \right].
\end{align}
Note that, if we consider the limit $a\to0$, eq.~(\ref{eq:egbgKerr}) reproduces the result
of ref.~\cite{Dvo06}, where neutrino spin oscillations in the
Schwarzschild metric were studied.

On the basis of eqs.~(\ref{eq:vierbKerr}) and~(\ref{eq:vectpot})
we get the nonzero components of electromagnetic field in the vierbein
frame,
\begin{equation}\label{eq:ebKerr}
  e_{1} = \frac{B_{0}ar_{g}}{2r^{2}\sqrt{\Xi_{0}}}(r^{2}-a^{2}),
  \quad
  b_{2} = -\frac{B_{0}}{2r^{2}}\sqrt{\frac{\Delta}{\Xi_{0}}}(2r^{3}+a^{2}r_{g}).
\end{equation}
It should be noted that, if we study a rotating BH ($a\neq0$), there
is a nonzero electric field in the system (see also ref.~\cite{Wal74}).

To find the explicit expressions for $U^{0}$ and $U^{\phi}$ we notice
that in the circular neutrino motion $u^{1}=u^{2}=0$ in eq.~(\ref{eq:uKerr}),
i.e. a neutrino moves along the straight line in the vierbein frame.
Using the fact that $u_{a}u^{a}=1$ and eq.~(\ref{eq:Uvierb}) we
get the following expressions for the remaining nonzero components
of $u^{a}$:
\begin{equation}\label{eq:u0u3eq}
  e_{g1}u^{0} = b_{g2}u^{3},
  \quad
  (u^{0})^{2}-(u^{3})^{2} = 1,
\end{equation}
which completely define the neutrino four velocity.

Finally, using eqs.~(\ref{eq:egbgKerr}) and~(\ref{eq:u0u3eq}),
we get that
\begin{equation}\label{eq:U0UphiKerr}
  U^{0} = \frac{\sqrt{2}x^{3/2}\pm\alpha}{\sqrt{2x^{3}-3x^{2}\pm2\sqrt{2}\alpha x^{3/2}}},
  \quad
  U^{\phi} = \pm\frac{1}{r_{g}}\frac{1}{\sqrt{2x^{3}-3x^{2}\pm2\sqrt{2}\alpha x^{3/2}}},
\end{equation}
where $x=r/r_{g}$ and $\alpha=a/r_{g}$. The upper sign refers to
the direct orbits, i.e. when a neutrino corotates with BH, and the
lower sign corresponds to retrograde ones (a particle counter-rotates).
Note that eq.~(\ref{eq:U0UphiKerr}) coincides with the analogous
expression obtained in ref.~\cite{BarPreTeu72} on the basis of the direct
solution of the geodesic eq.~(\ref{eq:Trajcurvedst}). We also mention that the more general description of the test particles dynamics in the Kerr metric is given in ref.~\cite{Dym86}. That analysis covers the motion of both massive and massless test particles on a wide range of trajectories which include not only equatorial circular orbits.

Basing on eq.~(\ref{eq:nuspinrot}) we can reformulate the neutrino
spin dynamics in the form of the effective Schr\"{o}dinger equation,
\begin{equation}\label{eq:effSchrod}
  \mathrm{i}\frac{\mathrm{d}\nu}{\mathrm{d}t} = H_{\mathrm{eff}}\nu,
\end{equation}
where the components of the wave function $\nu^{\mathrm{T}}=(\nu_{+},\nu_{-})$
describe different neutrino helicity states: $\nu_{+}$ corresponds to
$\bm{\zeta}$ parallel to $\mathbf{u}$ and $\nu_{-}$ stays for $\bm{\zeta}$
antiparallel to $\mathbf{u}$. The effective Hamiltonian in eq.~(\ref{eq:effSchrod})
has the form, $H_{\mathrm{eff}}=-(\bm{\sigma\Omega})$, where $\bm{\sigma}=(\sigma_{1},\sigma_{2},\sigma_{3})$
are the Pauli matrices and $\bm{\Omega}$ is the vector with the nonzero
components,
\begin{align}
  \Omega_{2} & =
  \frac{1}{2\gamma r_{g}}
  \left\{
    \mp\frac{1}{\sqrt{2}x^{3/2}} -
    \mu B_{0}r_{g}
    \frac{2\sqrt{2}x^{2}(x-1)\pm\alpha\sqrt{x}(2x-1)+\sqrt{2}\alpha^{2}}
    {x^{3/2}\sqrt{2x^{3}-3x^{2}\pm2\sqrt{2}\alpha x^{3/2}}}
  \right\},
  \nonumber \\
  \Omega_{3} & =
  \frac{G_{\mathrm{F}}n_{\mathrm{eff}}}{\sqrt{2}\gamma}
  \frac{\pm U_{f}^{0}-r_{g}U_{f}^{\phi}(\sqrt{2}x^{3/2}\pm\alpha)}
  {\sqrt{2x^{3}-3x^{2}\pm2\sqrt{2}\alpha x^{3/2}}}.
  \label{eq:Omega23}
\end{align}
To derive eq.~(\ref{eq:Omega23}) we use eqs.~(\ref{eq:uKerr})-(\ref{eq:U0UphiKerr}).

Supposing that $B_{0}$ and $n_{\mathrm{eff}}$ do not depend on time,
we can solve eq.~(\ref{eq:effSchrod}) analytically. Assuming that
initially only left-polarized neutrinos are present, i.e. $\nu^{\mathrm{T}}(0)=(0,1)$,
we get the probability to detect a right-polarized neutrino in the
form,
\begin{equation}\label{eq:Ptr}
  P(t) = P_{\mathrm{max}}\sin^{2}
  \left(
    \frac{\pi}{L_{\mathrm{osc}}}t
  \right),
\end{equation}
where
\begin{equation}\label{eq:PmaxLosc}
  P_{\mathrm{max}} = \frac{\Omega_{2}^{2}}{\Omega_{2}^{2}+\Omega_{3}^{2}},
  \quad
  L_{\mathrm{osc}} = \frac{\pi}{\sqrt{\Omega_{2}^{2}+\Omega_{3}^{2}}},
\end{equation}
are the amplitude and the oscillations length.

Now we can compare our results with the findings of previous works
where neutrino oscillations in gravitational backgrounds were studied.
Neutrino spin oscillations in the combination of the gravitational
field of a nonrotating BH, described by the Schwarzschild metric,
and a homogeneous magnetic field were studied in ref.~\cite{SorZil07}.
Setting $a=0$ and $n_{\mathrm{eff}}=0$ in eqs.~(\ref{eq:Omega23})-(\ref{eq:PmaxLosc})
we reproduce the result of that work.

Let us consider another class of the neutrino trajectories which imply the particle propagation from a region close to BH towards infinity. We shall assume that initially neutrinos are emitted radially. Note that, because of the symmetry reasons, a purely radial trajectory is allowed only if a particle propagates along the rotation axis of BH. However, if one studies an UHE particle, the deflection angle is small (see section~\ref{sec:APPL} for the estimates). Therefore we can assume that $U^\phi \approx 0$. As previously we shall study the equatorial neutrino motion, i.e. $U^\theta = 0$ and $\theta = \pi/2$.

Using eqs.~\eqref{eq:Kerrmetr}-\eqref{eq:vierbKerr} we get the new components of the neutrino four velocity in the vierbein frame,
\begin{equation}\label{eq:uKerrrad}
  u^{a} =
  \left(
    rU^{0}\sqrt{\frac{\Delta}{\Xi_{0}}},\frac{rU^r}{\sqrt{\Delta}},0,-\frac{ar_{g}U^{0}}{\sqrt{\Xi_{0}}}
  \right),
\end{equation}
and the nonzero components of vectors $\mathbf{e}_g$ and $\mathbf{b}_g$,
\begin{equation}\label{eq:egbgKerrrad}
  e_{g1} =
  -\frac{U^0}{2r^{2}\sqrt{\Xi_{0}}}
  (r^{2}+a^{2}),
  \quad
  e_{g3} =
  -\frac{a r_g U^r}{2 \sqrt{\Delta} \Xi_{0}}
  (a^{2} + 3 r^2),
  \quad
  b_{g2} =
  -\frac{r_g a U^0}{2r^{2}}\sqrt{\frac{\Delta}{\Xi_{0}}}.
\end{equation}
The background matter and external electromagnetic field are supposed to have the same configurations (see eqs.~\eqref{eq:gKerr} and~\eqref{eq:ebKerr}).

On the basis of eqs.~\eqref{eq:ebKerr}, \eqref{eq:uKerrrad}, and~\eqref{eq:egbgKerrrad} we derive the components of the vector $\bm{\Omega}$, cf. eq.~\eqref{eq:Omega23}, as
\begin{align}
  \Omega_{1} = &
  \frac{G_\mathrm{F}}{\sqrt{2}\gamma}
  \frac{rU^r}{\sqrt{\Delta}}
  \left\{
    g^0 + g^3 \frac{a r_g U^0}{\sqrt{\Xi_0} + r U^0 \sqrt{\Delta}}
  \right\},
  \notag
  \\
  \Omega_{2} = &
  \frac{\mu B_0}{2 r^2 \Xi_0}
  \left[
    (r^2 - a^2) a^2 r_g^2 - \Delta r (2r^3 + r_g a^2)
  \right]
  \notag
  \\
  & -
  \frac{a}{4\gamma}
  \bigg\{
    \frac{r_g U^0}{r^2} \sqrt{\frac{\Delta}{\Xi_0}} +
    \frac{r_g}{\Xi_0}
    \left(
      1 + r U^0 \sqrt{\frac{\Delta}{\Xi_0}}
    \right)^{-1}
    \notag
    \\
    & \times
    \left[
      \frac{(U^{0})^2 r_g}{r^2} (r^2 + a^2) -
      \frac{(U^{r})^2 r}{\Delta^2} (a^2 + 3 r^2)
    \right]
  \bigg\},
  \nonumber \\
  \Omega_{3} = &
  - \frac{G_\mathrm{F}}{\sqrt{2}\gamma}
  \left\{
    g^0 \frac{a r_g U^0}{\sqrt{\Xi_0}} +
    g^3
    \left[
      1 + \frac{(a r_g U^0)^2}{\Xi_0 + r U^0 \sqrt{\Delta\Xi_0}}
    \right]
  \right\},
  \label{eq:Omega23rad}
\end{align}
where $g^0$ and $g^3$ and given in eq.~\eqref{eq:gKerr}. Note that one has to add the law of motion to eq.~\eqref{eq:Omega23rad}. On the basis of eq.~\eqref{eq:Kerrmetr} and supposing that neutrinos are ultrarelativistic, i.e. $\mathrm{d}\tau \approx 0$, as well as $\theta = \pi/2$ and $\mathrm{d}\phi \approx 0$, we obtain that
\begin{equation}\label{eq:lmrad}
  \frac{\mathrm{d}r}{\mathrm{d}t} \approx
  \left(
    1 - \frac{r_g}{r}
  \right)
  \sqrt{
    1 + \frac{a^2}{r^2-r_g r}
  }.
\end{equation}
Using this expression one gets the relation between $U^0$ and $U^r$.

Let us rewrite eq.~\eqref{eq:Omega23rad} for UHE neutrinos, which have $U^0 \gg 1$. Expanding eqs.~\eqref{eq:Omega23rad} and~\eqref{eq:lmrad} in this limit we get
\begin{align}
  \Omega_{1} = &
  \frac{G_\mathrm{F} n_\mathrm{eff}}{\sqrt{2}}
  \left[
    U_f^0
    \left(
      1 - \frac{1}{x}
    \right) +
    r_g U_f^\phi \frac{\alpha}{x}
  \right],
  \notag
  \\
  \Omega_{2} = &
  - \mu B_0
  \left(
    1 - \frac{1}{x}
  \right) +
  \frac{\alpha}{4 r_g}
  \frac{2x - 3}{x^{7/2}\sqrt{x-1}} ,
  \nonumber \\
  \Omega_{3} = &
  - \frac{G_\mathrm{F} n_\mathrm{eff}}{\sqrt{2}}
  \frac{\alpha}{x^2}
  \sqrt{1-\frac{1}{x}}
  U_f^0,
  \label{eq:Omega23radf}
\end{align}
where the parameters $x$ and $\alpha$ are defined after eq.~\eqref{eq:U0UphiKerr}. Note that in eq.~\eqref{eq:Omega23radf} we keep terms up to linear in $\alpha$. In the next section we shall discuss some applications of our results.

\section{Applications\label{sec:APPL}}

In this section we discuss some applications of the results obtained
in section~\ref{sec:KERR}. We shall study both circular orbits and the radial neutrino propagation. In particular, we consider neutrino oscillations on a circular orbit
very close to BH, when the dynamics of the neutrino spin is determined
mainly by the gravity interaction. We compare our results with the
findings of previous works. Then we discuss the effect of the suppression of spin oscillations in case the neutrino interactions with gravitational and electromagnetic fields have the comparable strengths. Finally we consider the evolution of low
energy neutrinos when both gravity, matter, and magnetic interactions
equally contribute the dynamics of spin oscillations. Concerning radially propagating UHE neutrinos, first we shall demonstrate that the direct contribution of gravity to spin oscillations is small. Then we shall analyze the suppression of neutrino oscillations in a strong magnetic field and a relativistic accretion disk around a Kerr BH.

First let us discuss a neutrino orbit close to BH. We also suppose that the contribution of gravity to the neutrino spin evolution is dominant. Using eq.~\eqref{eq:Omega23}, we get that this approximation is valid if
\begin{equation}
  B_0 \ll 1.1 \times 10^{10} \times
  \left(
    \frac{\mu}{10^{-12}\thinspace\mu_\mathrm{B}}
  \right)^{-1}
  \left(
    \frac{M}{M_\odot}
  \right)^{-1}
  \text{G},
  \quad
  n_\mathrm{eff} \ll 7.4 \times 10^{26} \times
  \left(
    \frac{M}{M_\odot}
  \right)^{-1}
  \text{cm}^{-3},
\end{equation}
where $\mu_{\mathrm{B}}$ is the
Bohr magneton and $M_{\odot}=2\times10^{33}\thinspace\mathrm{g}$
is the solar mass.

Note that, in the considered
case, oscillations are in resonance since $P_{\mathrm{max}}\approx1$.
Using eq.~(\ref{eq:PmaxLosc}) we get that
\begin{equation}\label{eq:Loscgrav}
  L_{\mathrm{osc}} = 2\pi r_{g}
  \frac{2x^{3}\pm\sqrt{2}\alpha x^{3/2}}{\sqrt{2x^{3}-3x^{2}\pm2\sqrt{2}\alpha x^{3/2}}}.
\end{equation}
It is interesting to compare eq.~(\ref{eq:Loscgrav}) with the analogous
expression found in ref.~\cite{AlaHos13}, where neutrino spin oscillations
in Kerr background were studied. There is a significant discrepancy
of $L_{\mathrm{osc}}$ in eq.~(\ref{eq:Loscgrav}) and $L_{\mathrm{osc}}$
found in ref.~\cite{AlaHos13} for orbits corresponding to small $x\sim1$.
Nevertheless the asymptotic value of $L_{\mathrm{osc}}$ at $x\gg1$,
derived in our work, coincides with the result of ref.~\cite{AlaHos13}.
Such a discrepancy of the results may be explained by the fact that
the authors of ref.~\cite{AlaHos13} seemed to adopt the incorrect law
of motion of a neutrino on a circular orbit, cf. ref.~\cite{BarPreTeu72}.
Moreover, apparently only counter-rotating neutrinos were taken into
account in ref.~\cite{AlaHos13}.

\begin{figure}
  \centering
  \includegraphics[scale=0.3]{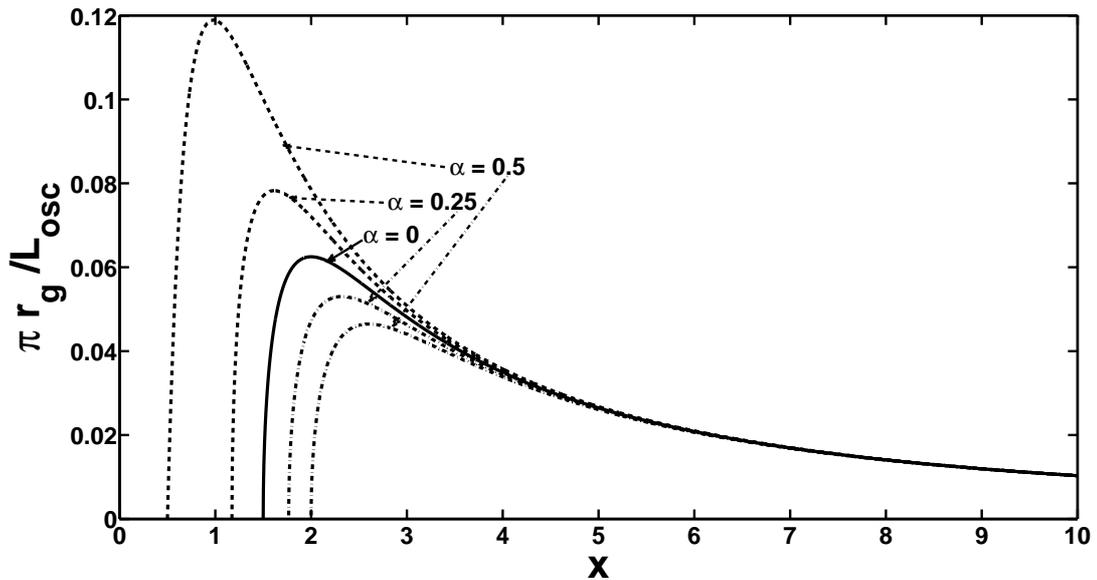}
  \caption{
  The function $\pi r_{g}/L_{\mathrm{osc}}=|\Omega_{2}|r_{g}$ versus
  $x=r/r_{g}$ for $\alpha=0$, $0.25$, and $0.5$. The case $\alpha=0$
  (solid line) corresponds to a nonrotating BH, described by the Schwarzschild
  metric, previously studied in ref.~\cite{Dvo06}. For $\alpha\neq0$,
  dashed lines are depicted for direct orbits and dash-dotted lines
  for retrograde ones.}
  \label{fig:Omega2vsx}
\end{figure}

In figure~\ref{fig:Omega2vsx} we present the dependence $\pi r_{g}/L_{\mathrm{osc}}=|\Omega_{2}|r_{g}$
versus $x=r/r_{g}$ for different values of $\alpha$. It can be seen
that for retrograde orbits the maximal value of $|\Omega_{2}|$ decreases
with $\alpha$, whereas for direct orbits this value increases with
$\alpha$.

If we discuss the maximal possible angular momentum of BH corresponding
to $\alpha=1/2$, we get from figure~\ref{fig:Omega2vsx} that for
a direct neutrino orbit with $r\approx r_{g}$, the frequency of neutrino
spin oscillations reaches its maximal value $\Omega_{2}^{+}\approx0.12r_{g}^{-1}$.
For a retrograde orbit with $r\approx2.60r_{g}$ we get the maximal
frequency $\Omega_{2}^{-}\approx0.05r_{g}^{-1}$. Considering BH with
$M=10M_{\odot}$, we obtain the corresponding oscillations lengths,
$L_{\mathrm{osc}}^{+}\approx7.85\times10^{7}\thinspace\text{cm}$
and $L_{\mathrm{osc}}^{-}\approx1.88\times10^{8}\thinspace\text{cm}$. On the basis of these oscillations lengths one gets the times of the neutrino spin-flip, $T_\mathrm{sf} = L_{\mathrm{osc}}/2$, as $T_\mathrm{sf}^{+} \approx 1.31 \times 10^{-3}\thinspace\text{s}$
and $T_\mathrm{sf}^{-} \approx 3.13 \times 10^{-3}\thinspace\text{s}$.

Note that for $\alpha=1/2$ the minimal possible radius of a stable
retrograde orbit is equal to $r_{\mathrm{min}}=4.5r_{g}$~\cite{BarPreTeu72}.
Therefore the obtained $\Omega_{2}^{-}$ corresponds to an unstable
orbit. In case of a direct orbit $r_{\mathrm{min}}\to r_{g}/2$. In
figure~\ref{fig:Omega2vsx} we get that in this case $\Omega_{2}=0$.

\begin{figure}
  \centering
  \includegraphics[scale=0.3]{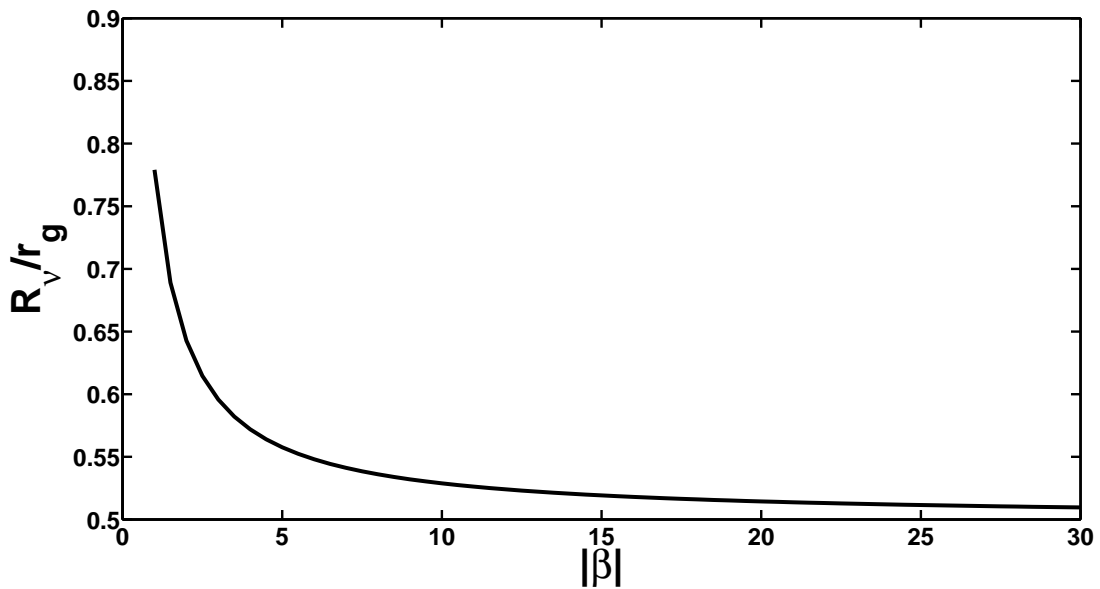}
  \caption{
  The radius of the neutrino orbit $R_\nu$ corresponding to the suppression of
  neutrino spin oscillations ($\Omega_2 = 0$) versus the absolute value of
  $\beta = \mu B_0 r_g$. This plot was built for a direct neutrino orbit,
  $\mu<0$, and $\alpha = 1/2$.}
  \label{fig:suppression}
\end{figure}

Now let us consider the neutrino spin evolution under the influence of both gravitational and external magnetic fields. Using eqs.~\eqref{eq:Omega23}-\eqref{eq:PmaxLosc}, one finds that at certain strength of the magnetic field, neutrino spin oscillations can be suppressed. It happens when $\Omega_2 = 0$. Of course, here we assume that the density of background matter is small but nonvanishing for the total Hamiltonian to be nonzero. Let us discuss the case when a neutrino corotates with BH and $\mu < 0$. In figure~\ref{fig:suppression} we show such a radius of the neutrino orbit $R_\nu$, at which $\Omega_2 = 0$, versus the absolute value of the parameter $\beta = \mu B_0 r_g$. Here we consider a particular case corresponding to $\alpha = 1/2$.

To get some estimates, we take that the BH mass $M=10M_{\odot}$ and the neutrino magnetic moment $|\mu| = 10^{-12}\mu_{\mathrm{B}}$. This value of $\mu$ is consistent with the astrophysical upper bound for the magnetic moment of a Dirac neutrino derived in ref.~\cite{KuzMik07}. Using figure~\ref{fig:suppression}, we obtain that, at $B_0 \sim (10^{9} - 10^{10}) \thinspace \text{G}$, spin oscillations can be suppressed for a neutrino moving close to BH on the orbit with $R_\nu \sim r_g$. It should be noted that for other types of the neutrino motion, e.g., corresponding to retrograde orbits, $\Omega_2$ always differs from zero, i.e. the described suppression does not take place.

Let us consider the situation when the matter contribution to the dynamics of neutrino oscillations is significant. At this moment we have to specify the law of motion of background fermions in an accretion disk. We discuss the situation when the motion of background fermions is affected mainly by the gravity. It can happen if the magnetic field in the system is weak. Thus, the fermions should move along geodesic lines. Using eq.~\eqref{eq:Omega23} and the analog of eq.~\eqref{eq:U0UphiKerr} for $U_f^0$ and $U_f^\phi$, one gets that the resonance in neutrino oscillations (i.e., the situation when $\Omega_{3} \approx 0$) can take place when the accretion disk and neutrinos rotate in the same direction. If this situation is implemented, neutrinos and background fermions are at rest with respect to each other. In other cases neutrino oscillations are suppressed.

Now we discuss the case when the contributions to the dynamics of spin oscillations from both matter, a magnetic field, and the gravity are of the similar size. In this situation we
have to study large neutrino orbits. Note that background
fermions, which can be electrically charged, should be also involved
in the electromagnetic interaction. Thus
the motion of the fermions can be far from geodesics. Therefore we have to account for
their law of motion relying on a phenomenological model of an accretion
disk. We can suppose that background matter is nonrelativistic and
rotates with the constant angular velocity, i.e. $U_{f}^{\mu}=(1,0,0,\omega_{f})$.
Other models of accretion disks are described in the recent review~\cite{AbrFra13}.

At large distance from BH the components of the vector $\bm{\Omega}$
have the form,
\begin{align}\label{eq:Omega23br}
  \Omega_{2} = &
  \mp\frac{1}{2\sqrt{2}r_{g}x^{3/2}}-\mu B_{0}+\dotsb,
  \notag
  \\
  \Omega_{3} = &
  -\frac{G_{\mathrm{F}}n_{\mathrm{eff}}}{\sqrt{2}}
  \left\{
    r_{g}\omega_{f}\mp\frac{1}{\sqrt{2}x^{3/2}}
    \left[
      1+r_{g}\omega_{f}(1-\alpha)
    \right]
  \right\} +
  \dotsb,
\end{align}
Using eqs.~(\ref{eq:PmaxLosc}) and~(\ref{eq:Omega23br}) we get
that the resonance in neutrino oscillations is achieved when a particle
moves on the direct orbit with
\begin{equation}\label{eq:rescondnr}
  R_\nu \approx 0.79r_{g}^{1/3} \omega_{f}^{-2/3}
  \left[
    1 + r_g \omega_f (1 - \alpha)
  \right]^{2/3}.
\end{equation}
In this case we obtain the following expression for the oscillations length:
\begin{equation}\label{eq:Losclargeorb}
  L_{\mathrm{osc}}=\pi
  \left|
    \mu B_{0} + \frac{\omega_{f}}{2}
    \left[
      1 + r_g \omega_f (1 - \alpha)
    \right]
  \right|^{-1},
\end{equation}
on the basis of the general expression~\eqref{eq:PmaxLosc}.

Let us consider BH with $M = 10 M_{\odot}$ surrounded by an accretion
disk with the inner $R_{\mathrm{in}} \approx 1.6 R_{\odot}$ and outer $R_{\mathrm{out}} \approx 8 R_{\odot}$ radii (see,
e.g., ref.~\cite{PerBlu10}), where $R_{\odot}=7\times10^{10}\thinspace\mathrm{cm}$
is the solar radius. Suppose that a low energy neutrino moves inside this accretion disk on the orbit with $R_\nu = 5 R_{\odot}$. We also assume
that accretion disk rotates with $\omega_{f} \approx 2 \times 10^{-8} r_{g}^{-1}$.
In this case background matter is nonrelativistic since its velocity
does not exceed $\omega_f R_{\mathrm{out}} \approx 3.7 \times 10^{-3}$. For such a parameters of an accretion disk, the
resonance condition in eq.~(\ref{eq:rescondnr}) is fulfilled. It should be noted that, for the chosen parameters, the BH rotation weakly affects the resonance condition and the expression of the oscillations length in eqs.~\eqref{eq:rescondnr} and~\eqref{eq:Losclargeorb}.

Analogously to the previous estimate, we suppose that $\mu = 10^{-12}\mu_{\mathrm{B}}$. We also assume that the magnetic field on the neutrino trajectory is
$B_{0} = 10\thinspace\text{G}$. It means that in the vicinity of
BH the magnetic field should reach $B_{0}(R_{\nu}/r_{g})^{3} \approx 1.6 \times 10^{16}\thinspace\text{G}$. For such parameters we get that $\mu B_{0} \approx \omega_{f}/2$, i.e. magnetic interaction can also influence neutrino spin oscillations.

Note that magnetic fields $\sim 10^{16}\thinspace\text{G}$ can be present in some compact astrophysical objects (see, e.g., ref.~\cite{HeyKul98}). Using eq.~\eqref{eq:Losclargeorb} we get the typical time for the neutrino spin-flip $T_\mathrm{sf} \approx 1.3 \times 10^{4}\thinspace\text{s} \sim 10^{-3}\thinspace\text{yr}$ for the chosen parameters. This time scale of neutrino oscillations is much shorter than characteristic time of the strong magnetic field decay found in ref.~\cite{HeyKul98}. Thus, neutrinos will experience multiple transitions between active and sterile states.

The most probable application of spin oscillations of neutrinos moving on closed orbits around BH is the description of low energy particles which can gravitationally cluster around massive astrophysical objects. Note that massive sterile neutrinos can contribute to warm dark matter~\cite{LesPas12}. Neutrinos can also interact with new bosonic dark matter particles. Although the form of this interaction is still unclear, there are some phenomenological models which involve only left-handed neutrino chiral projections (see, e.g., ref.~\cite{Man06}). Therefore, if there are multiple transitions between left- and right-polarized states under the influence of gravitational and magnetic fields as well as background matter, as described in our work, the rates of interaction of such neutrinos with bosonic dark matter particles can be up to two times reduced. It can have an implication for the calculation of the mean free path of these neutrinos~\cite{Man06}.

As we have seen, the effects related to spin oscillations of neutrinos moving on closed orbits around BH are rather difficult to observe. Thus, let us study the situation when particles escape the BH region. The analytical description of spin oscillations is possible for UHE neutrinos since we can neglect the deflection of the trajectory. We can evaluate the deflection angle as $\delta \phi \sim \alpha/(E_\nu r_g)$. Taking $M = 10 M_\odot$, $\alpha = 0.5$, and the neutrino energy $E_\nu = 10^9\thinspace\text{GeV}$, we get that $\delta \phi \sim 10^{-30} \ll 1$.

First, we shall consider the case when only the neutrino interaction with gravity is present. In this situation only $\Omega_2 \neq 0$ in eq.~\eqref{eq:Omega23radf}. The transition probability for $\nu_{+{}} \leftrightarrow \nu_{-{}}$ oscillations can be calculated as
\begin{equation}\label{eq:Ptrrad}
  P(x) =
  \sin^2(\Phi),
  \quad
  \Phi(x) =
  \int_{x_0}^x \mathrm{d} x'
  \frac{x'}{x'-1} r_g \Omega_2(x'),
\end{equation}
where the additional factor $x'/(x'-1)$ in the integrand is owing to the law of motion in eq.~\eqref{eq:lmrad}. To derive eq.~\eqref{eq:Ptrrad} we take into account that, in the linear approximation in $\alpha$, the neutrino trajectory in the vierbein frame is a straight line along the first axis.

\begin{figure}
  \centering
  \includegraphics[scale=0.5]{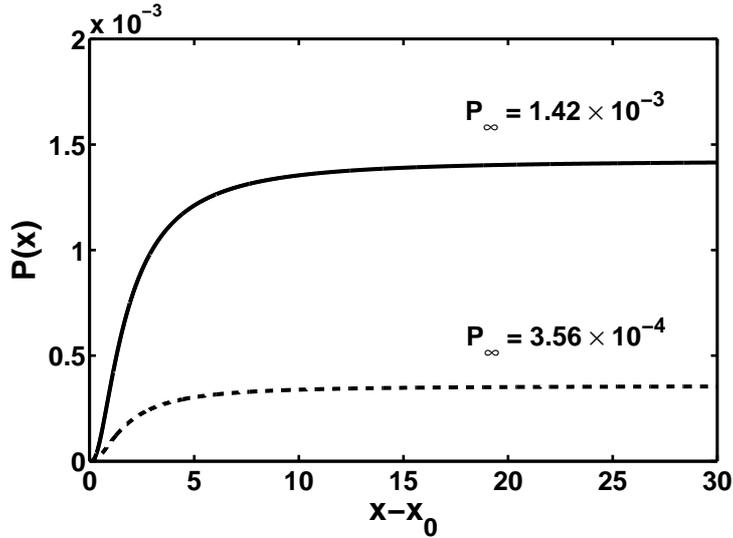}
  \caption{
  The transition probability for spin oscillations of radially moving neutrinos
  in the gravitational field of rotating BH. Here we take that $x_0 = 1.5$.
  The solid line corresponds to $\alpha = 0.5$ and the dashed line to
  $\alpha = 0.25$. The quantities $P_\infty$ stay for the asymptotic
  values of the transition probability at $x \gg x_0$.}
  \label{fig:Ptrvsxrad}
\end{figure}

The function $P(x)$ for $x_0 = 1.5$ is shown in figure~\ref{fig:Ptrvsxrad}. Although the motion of a test particle in the Kerr metric is allowed even for smaller $x$, one should not choose the values of the initial radial coordinate in the range $0.5 < x_0 < 1.5$ since the deflection of the trajectory can be significant. One can see in figure~\ref{fig:Ptrvsxrad} that the maximal transition probability for $\alpha = 0.5$ is $\sim 0.1\%$, that is small. Nevertheless, if one considers the neutrino emission from regions close the BH surface and accounts for the effects nonlinear in $\alpha$, the resulting transition probability can reach significant values. However, this situation requires a special analysis of exact solutions of the geodesic equation~\eqref{eq:Trajcurvedst} or~\eqref{eq:Uvierb}.

Note that the description of the evolution of UHE neutrinos in external fields can be important for explanation of the deficit of UHE neutrinos in cosmic rays reported in ref.~\cite{Abb12}. These UHE neutrinos can be emitted in gamma ray bursts or from active galactic nuclei, where both dense matter, strong magnetic and gravitational fields are present. Recently various possible solutions of the deficit of UHE neutrinos were proposed. Some of them are based on the attenuation of the neutrino flux because of the scattering off bosonic dark matter particles~\cite{Man06}, oscillations between active and hypothetical sterile neutrinos which are quasi-degenerate in masses~\cite{EmsFar12}, and the neutrino decay~\cite{BaeBusWin12}. A more plausible solution involves neutrino spin oscillations in strong magnetic fields~\cite{Bar12}. The authors of ref.~\cite{Bar12} used an important assumption of the small contribution of matter interaction to the neutrino spin evolution to obtaint the enhancement of the transition probability.

To analyze the neutrino spin precession in external fields more carefully, we shall consider UHE neutrinos produced in the vicinity of a rotating BH surrounded by an accretion disk. As we have seen, the direct gravity contribution to the dynamics of spin oscillations is small and can be omitted. Using eq.~\eqref{eq:Omega23radf} as well as the analog of linearized eq.~\eqref{eq:U0UphiKerr} for background fermions, we get the components of the vector $\bm{\Omega}$,
\begin{align}
  \Omega_{1} = &
  \frac{G_\mathrm{F} n_\mathrm{eff}}{\sqrt{2x^3 - 3x^2}}
  \left[
    x^{3/2}
    \left(
      1 - \frac{1}{x}
    \right) \mp
    \frac{\alpha x^2}{\sqrt{2}(2x^3 - 3x^2)}
  \right],
  \notag
  \\
  \Omega_{2} = & - \mu B_0
  \left(
    1 - \frac{1}{x}
  \right),
  \nonumber \\
  \Omega_{3} = &
  - G_\mathrm{F} n_\mathrm{eff}
  \frac{\alpha\sqrt{x-1}}{x\sqrt{2x^3 - 3x^2}},
  \label{eq:Omega23GRB}
\end{align}
where the upper sign corresponds to corotating BH and an accretion disk and the lower sign to counter-rotating ones. Taking into account eq.~\eqref{eq:lmrad}, one gets the effective Schr\"{o}dinger equation
\begin{equation}\label{eq:effSchGRB}
  \mathrm{i}\frac{\mathrm{d}\nu}{\mathrm{d}x} = H_\mathrm{eff} \nu,
  \quad
  H_\mathrm{eff} = - \frac{x}{x-1} r_g (\bm{\sigma}\bm{\Omega}).
\end{equation}
Since UHE neutrinos propagate along the first axis in the vierbein frame, the normalized wave functions corresponding to right- and left-polarized states read, $\nu_{\pm{}}^\mathrm{T} = (1, \pm 1)/\sqrt{2}$.

To analyze eq.~\eqref{eq:effSchGRB} we assume that $B_0(x) = 10^{12}\thinspace\text{G}/x^3$. This dependence corresponds to a dipole magnetic field. The accretion disk is supposed to consist of hydrogen plasma with the mass density $\rho = 10^2\thinspace\text{g}\cdot\text{cm}^{-3}$. Note that this value of the accretion disk density is rather moderate. Much denser disks are discussed in ref.~\cite{MacWoo99}. BH is taken to have $M = 10 M_\odot$. We also suppose that $\mu = 10^{-12}\mu_{\mathrm{B}}$ and consider counter-rotating BH and the accretion disk.

\begin{figure}
  \centering
  \includegraphics[scale=1]{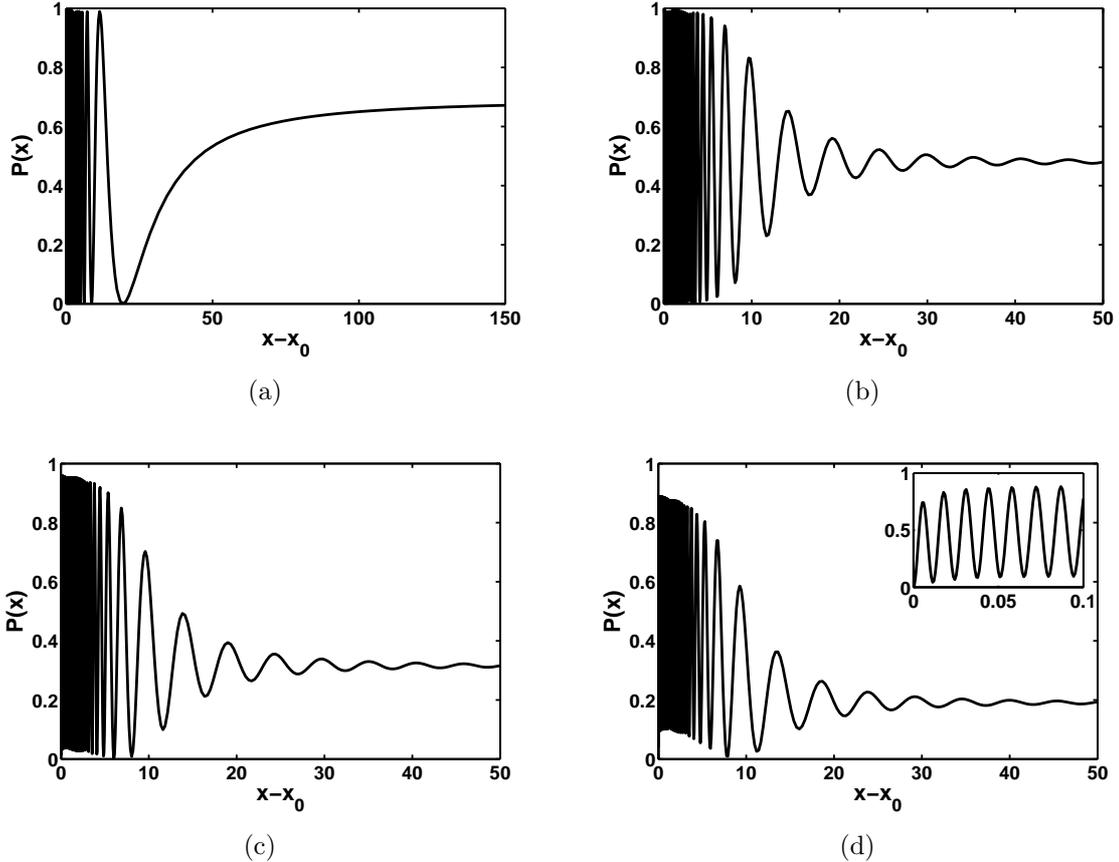}
  \caption{
  The transition probabilities for $\nu_{-{}} \leftrightarrow \nu_{+{}}$ oscillations versus the distance corresponding to the numerical
  solutions of eq.~\eqref{eq:effSchGRB}. (a) The case when only the magnetic interaction is present. (b-d) The situation when besides the magnetic field one has the neutrino interaction with an accretion disk; the panel~(b) corresponds to $\alpha = 0$ (Schwarzschild BH), the panel~(c) is plotted for $\alpha = 0.25$, and the panel~(d) for $\alpha = 0.5$. The inset in panel~(d) shows the behavior of $P(x)$ at small $x-x_0$.}
  \label{fig:PtrGRB}
\end{figure}

The corresponding transition probabilities are shown in figure~\ref{fig:PtrGRB}. First we notice that the ratio of magnetic and matter interactions $\mu B_0(1)/G_\mathrm{F} n_\mathrm{eff} \sim 10^3$. One may expect (as, e.g., in ref.~\cite{Bar12}) that the transition probability would be great since the matter interaction could be neglected. Indeed, one can see in figure~\ref{fig:PtrGRB}(b) that it is the case for a nonrotating BH since $P_\infty \sim 0.5$. As an example, we also present the transition probability for a purely magnetic neutrino interaction in figure~\ref{fig:PtrGRB}(a), for which one has $P_\infty \sim 0.7$. However, as one can see in figures~\ref{fig:PtrGRB}(c) and~\ref{fig:PtrGRB}(d), the transition probability can be significantly suppressed for a rotating BH. Therefore the statement of ref.~\cite{Bar12} that the amplitude of neutrino oscillations is almost equal to one, does not seem to be fully correct since the neutrino interaction with a relativistic accretion disk can significantly diminish the transition probability in some cases.

\section{Summary and discussion\label{sec:CONCL}}

In conclusion we mention that in the present work we have developed
the most general approach for the description of neutrino spin oscillations
in matter under the influence of gravitational and electromagnetic
fields. This approach is valid for the description of neutrino oscillations
in arbitrary gravitational and electromagnetic fields as well as in
background matter with arbitrary configuration of velocities and polarizations.

In section~\ref{sec:GENFORM} we have generalized the relativistic equation
for the description of the neutrino spin evolution in various external
fields. Previously the particle spin dynamics in an external field
was studied in refs.~\cite{DvoStu02,LobStu01,PomKhr98,BalKurZim02,Dvo06}. However
we have included to this equation the contribution of the neutrino
interaction with background fermions by means of the electroweak forces
when particles propagate in a curved space-time, cf. eq.~(\ref{eq:BMTcurvedst}).
It should be mentioned that the connection of electroweak and gravitational
interaction was also studied, e.g., in ref.~\cite{HehHeyKerNes76}. Then
we have rewritten this equation in the form, convenient for the description
of the neutrino polarization, see eqs.~(\ref{eq:nuspinrot}) and~(\ref{eq:vectG}).

It should be noted that in section~\ref{sec:GENFORM} we supposed that a spinning particle moves along a geodesic line in a curved space-time, cf. eq.~\eqref{eq:Trajcurvedst}. In general case this approximation is not valid. This fact was first mentioned in ref.~\cite{Pap51}. The analysis of the influence of the spin on the particle motion was further developed in ref.~\cite{RieHol93} using the pseudoclassical action in which the spin degrees of freedom are represented as Grassmann variables. Nevertheless, as was found in ref.~\cite{RieHol93}, in realistic cases, the deviation of the particle motion from geodesics is small. This problem was also analyzed in ref.~\cite{PomKhr98}, where it was stated that considering the parallel transport of the spin along the particle world line is the only possibility to construct an evolution equation linear in the spin. We also mention that the effects of the spin-orbit coupling can be sizable for a supermassive BH inside a dense star cluster (see, e.g., ref.~\cite{ThoHar85}).

To proceed, in section~\ref{sec:KERR}, we have discussed the neutrino
motion in the vicinity of a rotating BH surrounded by the accretion
disk and in the presence of an external magnetic field. Supposing that
a neutrino moves on a circular orbit, we have obtained the transition
probability for spin oscillations, see eqs.~(\ref{eq:Omega23})-(\ref{eq:PmaxLosc}).

Spin oscillations of a Dirac neutrino in the Schwarzschild background under the influence of an external magnetic field were previously studied in ref.~\cite{SorZil07} using the approach, developed in ref.~\cite{PapLam02} to account for the Mashhoon effect~\cite{Mas88} for a spin-$1/2$ particle in an external magnetic field. Choosing a suitable tetrade, the transition probability for neutrino oscillations was obtained in ref.~\cite{SorZil07} using the decomposition of the Dirac Hamiltonian. Note that the transition probability derived in ref.~\cite{SorZil07} for a circular neutrino orbit in the Schwarzschild metric is consistent with that obtained in our previous paper~\cite{Dvo06}.

The advantage of the description of neutrino spin oscillations based on the quasiclassical evolution eqs.~\eqref{eq:Uvierb} and~\eqref{eq:vectG} (see also refs.~\cite{PomKhr98,Dvo06}) compared to the approach of ref.~\cite{SorZil07} consists in the fact that one does not need to deal with a Dirac equation in a curved space-time. Within our formalism, the effective Hamiltonian, which is used for the description of neutrino oscillations, can be derived in a straightforward way if the components of a metric tensor are given.

Nevertheless, as we have mentioned in section~\ref{sec:KERR}, the results obtained in the present work are in agreement with that of ref.~\cite{SorZil07}, if we discuss a nonrotating BH and neglect the neutrino interaction with background matter.
However, since here we have described neutrino oscillations in more general backgrounds, our results should have wider applicability for the studies of the neutrino evolution in realistic astrophysical media such as
BHs in binary systems, quasars, and active galactic nuclei.

In section~\ref{sec:KERR}, we have also developed an approximate treatment of spin oscillations of UHE neutrinos supposing that these particles propagate radially. We have obtained the components of the vector $\bm{\Omega}$ which determine the effective Hamiltonian. This result is used in section~\ref{sec:APPL} to study spin oscillations of UHE neutrinos emitted in gamma ray bursts.

In section~\ref{sec:APPL} we have discussed some particular neutrino
trajectories. Firstly, we have studied neutrino oscillations on a circular orbit quite
close to BH, with the effect of gravity being dominant. We have obtained
the oscillations length, see eq.~(\ref{eq:Loscgrav}). Basing on
this expression we have corrected the results of ref.~\cite{AlaHos13},
where analogous expression was derived. We have also discussed the possibility of the suppression of neutrino oscillations when the contributions to the dynamics of the neutrino spin from gravity and electromagnetic field are comparable. Then we have studied neutrino
spin oscillations when a particle moves on the orbit with big radius.
In this case the contributions of all external fields are of the similar magnitude.
Considering the specific characteristics of BH and the accretion disk, we have
pointed out that our results may be applied for the description of the
spin dynamics of low energy neutrinos. In particular, the interaction of these neutrinos with bosonic dark matter particles, proposed in ref.~\cite{Man06}, can be significantly influenced if neutrinos experience multiple transitions between left- and right-polarized states.

In section~\ref{sec:APPL}, we have also studied the radial propagation of UHE neutrinos. It has been shown that the direct contribution of gravity to the dynamics of neutrino spin oscillations is small. However, we have demonstrated that the indirect contribution of gravitational interaction, which affects the motion of background fermions, can significantly change the behavior of the transition probability. We have discussed UHE neutrinos emitted close to a Kerr BH. These particles were taken to propagate through the relativistic accretion disk and interact with a strong magnetic field. In some situations the transition probability can be significantly suppressed. This result corrects the statement of ref.~\cite{Bar12} that the amplitude of the transition probability for spin oscillations of UHE neutrinos in a strong magnetic field is close to one.

\acknowledgments

I am thankful to V.~B.~Semikoz for helpful discussions and to FAPESP
(Brazil) for a grant.

\end{document}